# Direct Visualization of Two-State Dynamics on Metallic Glass Surfaces Well Below $T_g$

Sumit Ashtekar,[†] Gregory Scott,[†] Joseph Lyding[†,#] and Martin Gruebele[†,&*]

[†]Beckman Institute for Advanced Science and Technology, Department of Chemistry. [#]Department of Electrical and Computer Engineering, and [&]Department of Physics, University of Illinois, Urbana, IL 61801 USA.

**Abstract**

Direct atomically resolved observation of dynamics deep in the glassy regime has proved elusive for atomic and molecular glasses. Studies below the glass transition temperature $T_g$ are especially rare due to long waiting times required to observe dynamics. Here we directly visualize surface glass dynamics deep in the glassy regime. We analyze scanning tunneling microscopy movies of the surface of metallic glasses with time resolution as fast as 1 minute and extending up to 1,000 minutes. Rearrangements of surface cluster occur almost exclusively by two-state hopping ($P_{3\text{-state}} \approx 0.06$). All clusters are compact structures with a width of 2-8 atomic spacings along the surface plane. The two-state dynamics is both spatially and temporally heterogeneous. We estimate an average activation free energy of 14 $k_B T$ for surface clusters.

TOC Figure:

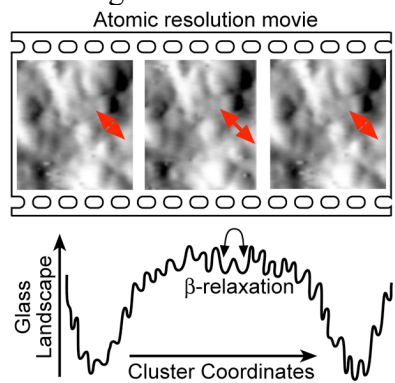

The glassy state of matter is of fundamental importance in chemistry and materials science. [1-4]. It is disordered yet cohesive, with a volume and free energy greater than the minimum possible at a given temperature. Glasses are rated on a fragile to strong scale[1], depending on how closely their relaxation obeys an Arrhenius temperature dependence. Well below the glass transition temperature $T_g$, cluster dynamics within the glass is a simple activated process with a distribution of rates, and corresponds to localized motions only (β-relaxation)[5,6]. Studies far below $T_g$ are rare.[7,8] As the glass transition temperature $T_g$ is approached, larger amplitude motions become feasible (α-relaxation) [9]. Heating above $T_g$ allows diffusive motions, so the liquid thermodynamic equilibrium state is reached[3].

Non-exponential relaxation dynamics is found in glass-forming systems and model colloids near the glass transition, revealing temporal heterogeneity.[2,10-12] It has proved difficult to completely disentangle homogeneous dynamics (temporal fluctuations of the rate at a single site) from heterogeneous dynamics (different rates for different sites).[13] NMR and other techniques have been used to infer the length scale of the moving clusters near $T_g$, yielding values ranging from 2-5 nm.[2,14] Theory predicts similar length scales for the dynamics on glass surfaces compared to the bulk, but with half the activation barriers[15]. Structural relaxation studies of polymer surfaces have yielded evidence for both increased and decreased relaxation rates at surfaces compared to the bulk[16,17].

**Results/Discussion**

To distinguish spatial and temporal heterogeneity, measure cluster size, determine the distribution of rates and the average activation barrier, we conducted an experiment directly visualizing glassy dynamics of a glass surface far below $T_g$. Using scanning tunneling microscopy movies, we studied metal alloys that are simple glass formers composed of a few size-mismatched atom types[18].

Figure 1 illustrates the data obtained and the basic data analysis. We used a home-built ultra-high vacuum scanning tunneling microscopy similar to one previously reported[19] (UHV-STM at ~$10^{-9}$ Pa) to acquire time-lapse images of the surfaces of three metallic glasses with atomic to near-atomic resolution. We studied Metglas 2605SA1 ($Fe_{78}B_{13}Si_8$)[20], Metglas 2705M ($Co_{69}Si_{12}B_{12}Fe_4Mo_2Ni_1$)[21] and Vitreloy1 ($Zr_{41.2}Ti_{13.8}Cu_{12.5}Ni_{10.0}Be_{22.5}$)[22] hereafter referred to as Fe-based, Co-based and Zr-based MG respectively. The glasses were degassed under UHV for 12 hours at 100 °C and sputtered for 1-2 hours under high vacuum with argon ions to remove surface oxides. An ion gun produced 1.5-2 keV argon ions in a chamber backfilled with high-purity argon gas to a pressure of 0.007 Pa. XPS measurements confirmed that the sputtering conditions were able to successfully clean the glass surface. The freshly sputtered glass samples were transferred from the preparation chamber directly to the UHV chamber.

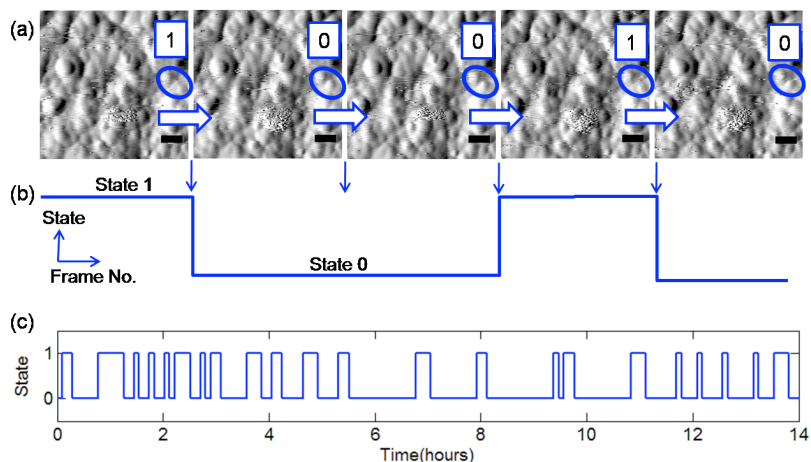

**Figure 1. Time-lapse images of Fe-based glass and single cluster trace (SCT) (a)** Five consecutive frames illustrating the two-state switching of a cluster of the surface of the Fe-based glass. The cluster switches between two positions 0.5 nm apart. The frames are separated by about 6 min. Spatial derivatives of the STM topographic images are shown **(b)** Construction of the SCT representing the two states of the cluster by "1" and "0" **(c)** full SCT of this cluster extending to about 14 hours. Scanning conditions: 2 V, 100 pA. Black scale bar: 5 Atomic Weighted Diameters (AWD) = 1.2 nm. A full movie is in SI.

UHV-STM scans were performed using electrochemically etched tungsten tips at 1-2 V bias voltage and 5-100 pA tunneling currents. High-resolution topographic images of the same area were collected successively, creating movies of the surface with 1 to 6 minute time resolution. Image registration compensated for small drift between successive images, no other processing was employed. Elevated temperature STM scans probed the temperature dependence of the observed surface kinetics at 80 and 150 °C. The glasses studied here have glass transition temperatures $T_g \gg$ 150 °C (Fe-based: 507 °C, Co-based: 520 °C, Zr-based: 352 °C[22]), so the studies conducted here all lie in the deep glassy regime. The metallic glasses we studied were prepared with very different critical cooling rates and have different fragilities (Fe-based and Co-based : $10^6$ K/s, m~110 = fragile; Zr-based: < 10 K/s, m~50 = strong)[23]. We were able to heat glass surfaces above $T_g$ and observe crystallization, confirming the very different morphology of the glass surface (see SI).

Figure 1 shows a series of two-state hopping events we observed (full raw data movie in SI Movie S1). An Fe-based MG cluster hops back and forth by ≈0.5 nm. The blue trace in Figure 1c shows the resulting single-cluster transitions (SCTs) between the two surface states as a function of time. No deposition of material from the tip was observed during the scans. Out of 50 separate moving clusters observed on 3 metal surfaces, only 4 did not undergo two-state transitions; three of these were sequential 3-state transitions and one cluster was observed to exhibit a 4-state transition. No diffusion was observed. Thus the probability that 3 free energy wells have comparable free energy and can interconvert with rates within our dynamic range (up to 200:1) is of the order $P\sim0.06$ compared to two-state dynamics. The glassy surface is essentially composed of two-state dynamical systems and immobilized clusters that do not move on a $10^3$ minutes time scale. The tunneling current between 5 and 100 pA, through local heating, has only a mild effect on accelerating hopping dynamics (see SI), so dynamics are observed 'natively' at low currents.

An interesting dynamical effect is observed in a few movies. The 'noise' that occurs in every image at the precise same location in Figure 1a (red arrow) is due to a cluster that hops rapidly while the STM scans over it, so the cluster cannot be resolved. Only three clusters were that fast. 50 were fast enough to hop multiple times while the STM scanned over them, and ≈22,000(these 22000 clusters include those which moved just once) were stationary in all the movies we scanned.

Figure 2 gives evidence for the spatial and temporal heterogeneity of the two-state transitions we observed on metallic glass surfaces. In figure 2a, clusters 1 (diameter ~5 Fe atoms) and 2 (diameter ~ 2 Fe atoms) are adjacent to one another on the same Fe-based glass surface. Cluster 1 has an equilibrium constant $K_{eq} = 0.9$, corresponding to a free energy difference of $\Delta G = 0.3$ kJ/mole between its two states. Its local free energy landscape corresponds to a nearly symmetric double well. Cluster 2 has $K_{eq} = 0.1$ and $\Delta G = 5.7$ kJ/mol, corresponding to an asymmetric double well. The relaxation rates for the two clusters are $k = <\tau_0^{-1}> + <\tau_1^{-1}> =$ 0.108 min$^{-1}$ and $k = 0.121$ min$^{-1}$(See Methods). The rate of cluster 1 temporarily slows down in the middle of the movie, and we observed such time-varying rates for 3 out of a total 50 clusters. Figure 2c shows an example of irreversible (within our dynamic range) 'aging' of a cluster, which switches from $k \approx 0.022$ min$^{-1}$ in the initial 4 hours of the movie to $k < 0.001$ min$^{-1}$ for the latter part of the movie (full Fig 2 movies are available in SI).

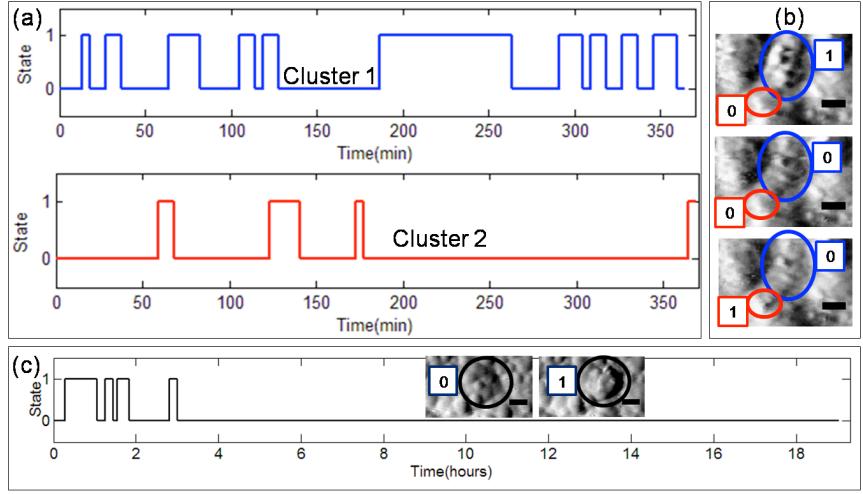

**Figure 2. Spatial and Temporal heterogeneities**. **(a)** Single cluster traces (SCTs) of a 'fast' cluster (blue) and a 'slow' cluster (red) from the same movie of the Fe-based glass. (Blue: $k$ = 0.108 min$^{-1}$) with ΔG= 0.27kJ/mole; red; $k$= 0.121min$^{-1}$ with ΔG = 5.72 kJ/mole.)(See Methods) Though separated by just 2 nm, the fast cluster switches 5 times more often than the slower one. **(b)** Consecutive images depicting the fast (blue) and slow (red) clusters of **(a)**, showing three combinations of "1" and "0" states. The time duration between the successive images is about 4.5 min. Scanning conditons: 1 V, 50 pA, **(c)** SCT of an aging cluster shows that it was active at first and stopped exhibiting movement in the later part of the movie. Inset depicts the "1" and "0" state of the cluster. The time duration between the successive images is about 12 min and 54 min. Scanning conditions: 2V, 100 pA. Scale bars: 5 Atomic Weighted Diameters (AWD) = 1.2 nm. Full movies are in SI.

Figure 3a summarizes the size distribution of the clusters observed in motion. The cluster diameter probability distribution decreases rapidly between 4 and 8 AWD (=atomic weighted diameters, see Methods). The paucity of clusters below 3 AWD may be an artifact of STM resolution (marked by the dotted black line). The 2-D cross sections in the surface plane are oval with aspect ratios ranging from 1:1 to 2.5:1 (see plot in SI). No strong correlation between 2-D cross section and rate exists, but an unknown fraction of each cluster is buried below the surface, so the apparent cluster size may differ from the actual size.

To estimate the average free energy barrier $\overline{\Delta G^{\dagger}}$ of the clusters observed in motion, we assumed a prefactor of $k_0$=1 ps$^{-1}$, and used the Arrhenius law

$$\bar{k} = k_0 e^{-\overline{\Delta G^{\dagger}}/RT} \quad (1)$$

together with our measured average rate $\bar{k}$ = (60 min)$^{-1}$ from Figure 3b. This yields $\overline{\Delta G^{\dagger}} \approx 14$ $k_B T_g$, less than half the activation energy estimated for the bulk (36 $k_B T_g$).[4] That the surface dynamics is faster than bulk dynamics agrees with results reported for surfaces of amorphous polystyrene [17]. Glass transition theory also predict a lowering of the surface activation energy by a factor of 2 from the bulk[15]. Arrhenius plots for individual clusters could not be measured because STM imaging and hence registration could not be maintained during heating with our current setup.

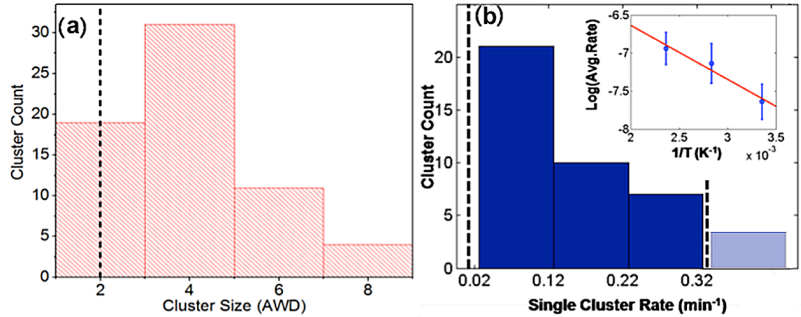

**Figure 3. Characteristics of hopping clusters. (a) Cluster size distribution** of the Fe-based glass at room temperature. This distribution decreases between 4 and 8 atomic weighted diameters (AWD), with an average at 4.1 AWD. The dotted line represents the average resolution of our STM scans, indicating that clusters below 2 AWD cannot be counted. 1 AWD = 0.24 nm **(b) Cluster rate distribution** of the Fe-based glass at room temperature. The rates were calculated from SCTs of 38 different clusters that hopped at least twice. The distribution decreases slowly as k increases. The vertical dotted lines showing the limits of minimum and maximum single cluster rate detectable by the experiments indicate the dynamic range for this study. The light blue bar corresponds to 3 clusters hopping faster than our time resolution (e.g. 'noise patch' in Figure 1). **Inset :Arrhenius plot:** The average cluster rate is in 2-state events per minute per $nm^2$ (including single switches). The slope of the linear fit indicates a low activation energy of 6 kJ/mol.

Figure 3b shows the rate coefficient distribution $P(k)$ for clusters that underwent at least two transitions, with dashed lines indicating our accessible dynamic range. The rate coefficients were calculated from the average dwell times in the two states as $k = <\tau_0^{-1}> + <\tau_1^{-1}>$ (see Methods). The distribution drops off slowly with increasing $k$. Also consistent with this broad rate distribution is the weak temperature dependence observed for the hopping rate per unit time and unit area averaged over all moving clusters (inset in Figure 3b). The slope for that Arrhenius plot is much lower than the activation barrier of individual clusters calculated above. The explanation is that a distribution of two-state barriers leads to a broad rate distribution. Thus slower clusters replace faster clusters in our observational time window when the sample is heated, keeping the rate averaged over many clusters nearly constant.

Figure 4 shows that two-state dynamics is not special to the Fe-based MG. The strong Zr-based MG and the Co-based MG, another more fragile glass, behave similarly (Figure 4a). All three glasses have similar average cluster sizes and widths (standard deviations) of the size distributions (Figure 4b). The Zr- and Co-based MG cluster shapes also mimic the Fe-based MG cluster shapes.

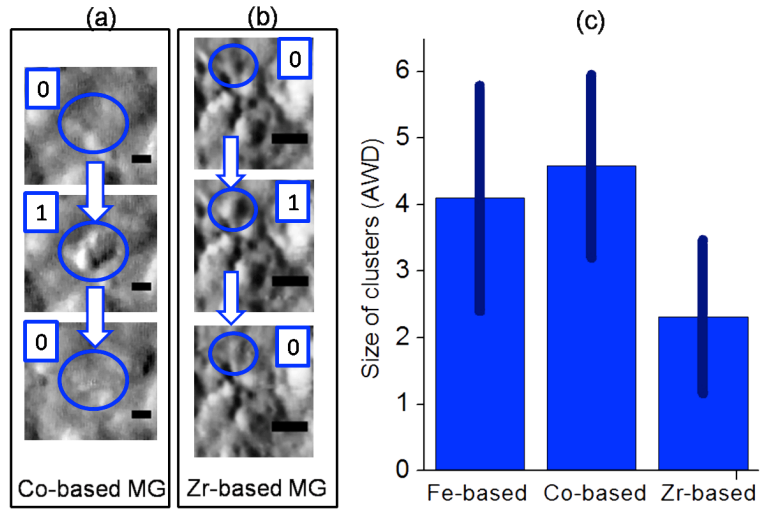

**Figure 4. Universality of the two-state dynamics in metallic glasses (a)-(b)** Two-state structural switching as observed in the Zr-based glass and Co-based glass. In **(a)**, the time duration between the successive images is about 5 min and 25 min. In **(b)**, the time duration between the successive images is about 6 min and 120 min. **(c)** The average cluster size of 3 different glass formers studied was found to be similar (dark blue bars), as was the standard deviation of the size distribution (dark blue bar). Scale bar: 5 weighted atomic diameters (AWD); 1.2 nm for Co-based, 1.3 nm for Zr-based glass

$\beta$ relaxations previously have been proposed to exist in metallic glasses[24-26], and localized random two-state motions with site heterogeneity and temporal heterogeneity are predicted by glass models[27,28]. We do observe a few 3- and 4-state clusters, but most of the motions we observe are indeed highly localized. Given the relatively high rate observed even well below $T_g$, these motions are probably best assigned as $\beta$ relaxations, although their projection at least in the x-y plane accessible to the STM is compact, and not especially elongated.

Random first order transition theory predicts a cluster dimension peaked at about 5 AWD for $\alpha$ relaxation, and slightly lower for $\beta$ relaxation of atomic glasses[4]. This prediction is in agreement with Figs. 3 and 4. The rate distribution and the small temperature dependence of the cluster-averaged rate in Figure 3b are also consistent with a predicted power law distribution[9], although our dynamic range is currently too limited to assign a specific functional form to $P(k)$.

In conclusion, STM movies directly visualize two-state dynamics (cooperative atomic motion) of clusters with a transverse size of 4-8 AWD on the surface of several atomic glasses well below their $T_g$. Three-state dynamics is rare ($P\approx0.06$), but its possible existence has been inferred previously from complex spectral trails in single molecule experiments.[29] Cluster motions are spatially heterogeneous, with different rates and free energy differences at different sites, even at adjacent sites. In addition, both intermittent rate fluctuations and irreversible (in our time window) 'aging' were observed for individual clusters. Such 'aging,' or cessation of dynamics, could be the atomic-level manifestation of macroscopic aging: slowly aged glasses settle into lower free energy minima, presumably resulting in reduced hopping dynamics.

**Experimental Methods**

**Glass preparation and characterization**    Metglas 2605 SA1 (1 mil thick foil) and Metglas 2705M (0.8 mil thick ribbon) were used as received from Metglas. Inc. These glasses were cut to size using ordinary scissors to fit the STM sample holders and were cleaned in acetone andisopropyl alchohol for 20 min each. The dull but smoother side of the foil for both these glasses was used for scanning rather than the shiny but rougher side.

Vitreloy 1 was received from LiquidMetal. Inc in 50x50mm plates with thickness 3.2 mm. These were cut to size to fit the sample holders using electric-discharge machining. The face of the sample to be scanned was polished using Buehler Gamma Micropolish with alumina particle size of 5 μm, 1 μm, 0.05 μm in succession.

X-ray photoelectron measurements were done on PHI5400 with a Mg source and equipped with differentially pumped ion gun. XPS scans were taken after brief periods of time of sputtering the glass surface to confirm the reduction and final disappearance of the Oxygen peak from the spectrum.

All glass samples were sputtered in the HV chamber (Base pressure: $4\times10^{-6}$Pa) and transferred to the attached UHV chamber for degassing. The degassed samples were then transferred to the UHV-STM chamber for scanning.

**Scanning tunneling microscopy**  All the STM images are presented in the spatial derivative mode to enable the readers to see the contrast between different clusters. Images are raw data with frames registered by shifting the (x,y) origin to maximize cross-correlation with a reference frame.

For temperature-dependence measurements, the sample was heated by a custom-designed sample attachment equipped with a resistive wire (Kanthal A1, 10 mil diameter) and a type K thermocouple to provide *in situ* temperature measurements accurate to ±3 °C. Power dissipation of 0.55 W and 1.53 W yielded sample temperatures of 80 °C and 150 °C, respectively.

**Data analysis** The atomic diameters for the three glasses were calculated by weighting their components to their atomic composition. The calculated atomic weighted diameters (AWD) for MGSA1 ($Fe_{78}B_{13}Si_8$), MG2705M ($Co_{69}Si_{12}B_{12}Fe_4Mo_2Ni_1$) and Vitreloy1 ($Zr_{41.2}Ti_{13.8}Cu_{12.5}Ni_{10.0}Be_{22.5}$) were 0.238 nm, 0.238 nm and 0.258 nm. The atomic radii for various components used for the calculations were Fe (0.125nm), B (0.09nm), Si (0.111nm), Co (0.126nm), Mo (0.139nm), Ni (0.121nm), Ti (0.136nm), Zr (0.148nm), Ni (0.121nm), Be (0.09nm).

The sizes of the rearranging clusters were determined by drawing a straight line in the spatial derivative image encompassing the ends of the clusters. For non-spherical clusters, an average of the two lines drawn at right angles was used as its size. Measured diameters appear slightly larger than actual diameters due to the convolution by the tip resolution. Therefore only the highest resolution scans were included for data analysis in this study.

Rates were computed using the dwell times $\tau_1$ in state 1 and $\tau_0$ in state 0, where the carets < > in the main text indicate averaging over all completely observed dwell events. The free energy difference was computed as $\Delta G = |RT\ln K_{eq}|$, and the equilibrium constant $K_{eq}$ was calculated as the ratio of the average dwell times in states 0 and 1.

The rates in Figure 3b were calculated by visually counting the number of rearranging clusters in a movie. That number was then divided by the area of the frames, and by the total time duration of the movie. To estimate the error bars for the average rate for a particular temperature, the clusters were assumed to follow Poisson distribution as to the number of clusters rearranging in each frame of the movie.

**Acknowledgement** We thank Craig Zeilenga for his help with the design of sample heating fixture and Scott Schmucker for his help with the construction of the sputtering chamber. We acknowledge financial support from the ACS Petroleum Research fund (AC-45421), and from the National Science Foundation (CHE-0948382). M.G. held the James R. Eiszner Chair while this work was carried out.

**Supporting Information available**: SI describes experiments in more detail, and provides three STM movies (from Figures 1 and 2) viewable with Quicktime. This material is available free of charge via the internet at http://pubs.acs.org.